%% file: 0_abs_ieee.tex
\documentclass[conference]{IEEEtran}
\IEEEoverridecommandlockouts

\usepackage{cite}
\usepackage{amsmath,amssymb,amsfonts}
\usepackage{algorithmic}
\usepackage{graphicx}
\usepackage{textcomp}
\usepackage{xcolor}

\usepackage{array}
\usepackage{multirow}
\usepackage{color}
\usepackage{algorithm}
\usepackage{tikz}
\usepackage{makecell}
\usetikzlibrary{shadows,arrows,shapes,positioning}
\usepackage{graphicx}
\usepackage{wasysym}
\newcolumntype{P}[1]{>{\centering\arraybackslash}p{#1}}
\newcolumntype{M}[1]{>{\centering\arraybackslash}m{#1}}
\usepackage{pifont}
\usepackage{enumitem}
\setlist{nolistsep}

\begin{document}

\title{Enhancing Robustness Against Adversarial Examples in Network Intrusion Detection Systems}

\author{\IEEEauthorblockN{Mohammad J. Hashemi}
\IEEEauthorblockA{\textit{Department of Computer Science} \\
\textit{University of Colorado Boulder}\\
Boulder, CO, USA \\
mohammad.hashemi@colorado.edu}
\and
\IEEEauthorblockN{Eric Keller}
\IEEEauthorblockA{\textit{Department of Electrical Computer and Energy Engineering} \\
\textit{University of Colorado Boulder}\\
Boulder, CO, USA \\
eric.keller@colorado.edu}
}

\maketitle

\newcommand{\blue}[1]{\textcolor{black}{#1}}

\begin{abstract}
The increase of cyber attacks in both the numbers and varieties in recent years demands to build a more sophisticated network intrusion detection system (NIDS). These NIDS perform better when they can monitor all the traffic traversing through the network like when being deployed on a Software-Defined Network (SDN). Because of the inability to detect zero-day attacks, signature-based NIDS which were traditionally used for detecting malicious traffic are beginning to get replaced by anomaly-based NIDS built on neural networks. However, recently it has been shown that such NIDS have their own drawback namely being vulnerable to the adversarial example attack. Moreover, they were mostly evaluated on the old datasets which don't represent the variety of attacks network systems might face these days. In this paper, we present Reconstruction from Partial Observation (RePO) as a new mechanism to build an NIDS with the help of denoising autoencoders capable of detecting different types of network attacks in a low false alert setting with an enhanced robustness against adversarial example attack. 
Our evaluation conducted on a dataset with a variety of network attacks shows denoising autoencoders can improve detection of malicious traffic by up to 29\% in a normal setting and by up to 45\% in an adversarial setting compared to other recently proposed anomaly detectors.
\end{abstract}

\begin{IEEEkeywords}
Intrusion Detection Systems, Neural Networks, Anomaly Detection, Adversarial Example
\end{IEEEkeywords}
\input{1_introduction.tex}
\input{2_background.tex}

\input{3_RePO.tex}

\input{4_eval}
\input{5_conclusion}

\bibliographystyle{IEEEtran}
\bibliography{sample-base}

\end{document}

%% file: 1_introduction.tex
\section{Introduction}
The continuous growth of network attacks in number, scale and complexity ~\cite{symantec_report} has caused a wide range of impacts to many individuals and exploited businesses from high monetary costs to more serious issues such as wide-scale power outages ~\cite{power_outage}. Because of the severe, adverse effects that such network attacks cause companies are expected to invest billions of dollars to find effective tools that detect and eliminate network intrusions~\cite{ids_for_cloud}.

\blue{ Network intrusion detection systems (NIDS) are one part in the line of defense against network attacks. Because of the technologies such as P4-based network telemetry, network function virtualization (NFV), cloud-native security services (such as what Zscaler provides) and network-wide view that the control plane in an SDN provides, these NIDS are getting more and more utilized to detect different types of threats  \cite{sdn_ids_survey,athena}.} But, signature-based NIDS which were traditionally being used to detect malicious traffic can't cope with today's variety of network attacks as they are incapable of detecting zero-day attacks which are significantly increased in recent years \cite{ids_survey2}.

In response, there has been a great deal of research and even commercial offerings which leverage machine learning (with deep neural networks) to augment the detection capabilities~\cite{deep_ids1,deep_ids2,deep_ids3,deep_ids4,kitsune,DAGMM,BiGAN,deep_ids_survey}.  These anomaly-based NIDS have been introduced due to their ability to detect zero-day attacks (for which there is no pre-existing signature) by looking for deviations from typical, benign network traffic. To do so, these NIDS are trained only on benign traffic. Then, during inference time, the NIDS measures how similar the new traffic is to the traffic seen during training time. Each packet or flow seen by the NIDS is given a similarity score and compared to a pre-defined threshold. If the packet or flow score exceeds the threshold, then the traffic is considered malicious. This threshold should be set in a way to make sure that the NIDS doesn't generate too many false alerts on benign traffic.

\blue{Recently, it has been shown that the detecting capability of ML-based NIDS including simple classifiers trained with supervision as well as complex anomaly-based NIDS trained in an unsupervised manner can be significantly reduced by the help of the adversarial example (evasion) attack \cite{adv_nids, sdn_adv}. This attack lets the attacker carefully and in many cases slightly manipulate malicious traffic to fool and bypass the NIDS while carrying out the original malicious intent without breaking the underlying network protocols. The deterministic behavior of the previously proposed anomaly-based NIDS makes it easy to craft adversarial examples against them. In addition, minimizing the reconstruction error of the benign traffic in the training phase used by some of the NIDS \cite{kitsune} based on the full observation of the inputs can lead to an over-generalization problem. It means that the model learns to reconstruct the malicious traffic which was not trained on, as good as benign traffic, making it hard for the model to distinguish between them leading to a poor detection rate.  Therefore, a new method for detecting malicious traffic is required to be more robust against adversarial example attack. Such a method should be able to detect a wide range of threats while generating a low number of false alerts. Because a high false alert rate significantly increases the required effort of security experts to manually sort through all alerts and differentiate true attacks from falsely identified attacks.}

\blue{In this paper, we present Reconstruction from Partial Observation (RePO) as a new method to build a more accurate NIDS in an unsupervised manner which is also more robust in the presence of adversarial examples by utilizing denoising autoencoders \cite{denoise_ae} and combining the inputs with multiple random masks before feeding them into the model. As we show, leveraging multiple random masks makes it harder to craft adversarial examples against the NIDS by making the model non-deterministic. Furthermore, it prevents the over-generalization problem we mentioned resulting in better distinguishment of malicious inputs from the benign traffic leading to a higher detection rate of the network attacks.}

In summary, we make the following contributions:
\begin{itemize}
    
    \item \blue{We leverage denoising autoencoders to build an NIDS that can detect malicious traffic better than previously proposed anomaly-based NIDS in a low false alert setting which is also more robust in an adversarial setting.}
    \item We show how a packet-based NIDS can be built with denoising autoencoders on top of raw values directly extracted from packet headers without any manual feature engineering. We also show how a flow-based NIDS can be built with them on top of manual features, calculated based on a whole flow.
    \item We evaluate our NIDS on a new network traffic dataset, which contains a wide range of attacks to show its effectiveness in detecting different types of attacks in both normal and adversarial setting.

\end{itemize}

%% file: 2_background.tex
\section{Background}
\subsection{Anomaly-based NIDS}

\begin{figure}
\centering
\includegraphics[width=85mm]{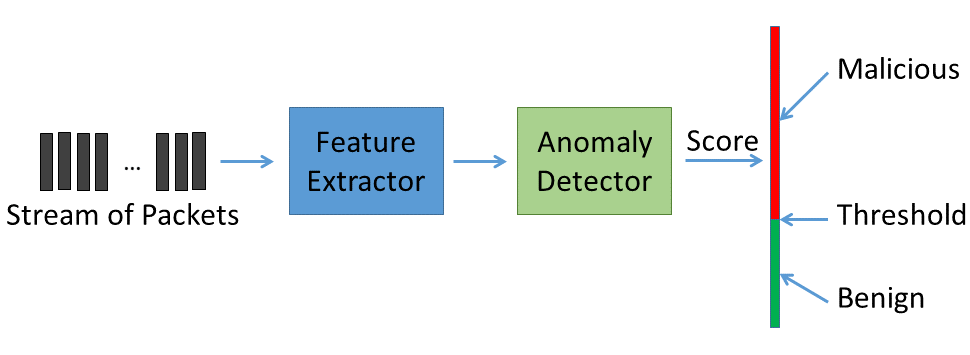}
\setlength{\belowcaptionskip}{-7pt}
\caption{Illustration of the structure of a typical anomaly-based NIDS.}
\label{fig:NIDS_structure}

\end{figure}

\blue{Anomaly-based NIDS can be built in many different ways. Figure \ref{fig:NIDS_structure} illustrates a typical anomaly-based NIDS. These NIDS have two major components: a feature extractor and an anomaly detector. In a nutshell, the feature extractor receives a stream of packets and extract features from them to feed them to the anomaly detector. Then, the anomaly detector outputs a score for each input it receives that gets compared against a threshold. If the score is less than the threshold the corresponding input will be predicted as benign otherwise it'll be predicted as malicious. The feature extractor can be built in two different ways. It can output a feature vector for each packet it receives. This feature vector is not based solely on the current packet; it can also consider the history of packets it has seen earlier. In this case, the features can be as simple as raw values extracted from packet headers or complex features created manually. We call the NIDS leverage such feature extractors packet-based NIDS. On the other hand, there are NIDS where their feature extractor builds a single feature vector with high-level features for a whole flow. We call these NIDS flow-based NIDS. The anomaly detector component can also be created in several different ways. Kitsune \cite{kitsune} which is a packet-based NIDS proposed to leverage an ensemble of autoencoders which are neural-networks trained to reconstruct a given input based on the full observation of that input. Then during the execution time, it calculates the final score for each input based on the reconstruction errors of the autoencoders. Zong et al. proposed DAGMM \cite{DAGMM} which trains two neural-network in an end-to-end fashion to calculates the energy of each flow in the Gaussian Mixture Model (GMM) framework and uses that energy to detect anomalies. Zenati et al. introduced a BiGAN-based approach \cite{BiGAN} that uses the reconstruction error of the generator and the output of the discriminator in the GAN (Generative Adversarial Network) framework to detect anomalies. We refer the readers to those papers for more details about each approach. We first show how one can build a packet-based NIDS with the RePO technique and compare it with Kitsune. Then we demonstrate that an anomaly detector built with RePO can also be utilized in the flow-based context and compare it with DAGMM and the BiGAN-based NIDS which were evaluated in such a setting. We chose these methods as our baselines as they were among the most highly cited NIDS which were published recently.}

\subsection{NIDS in Adversarial Setting}

It has been shown that deep neural networks used for tasks such as image classification \cite{biggio_evasion,intriguing}, speech-to-text systems \cite{carlini_audio}, face recognition \cite{sharif_face}, autonomous driving \cite{advsign}, malware detection \cite{grosse_malware}, etc. are vulnerable to adversarial example attacks (i.e. evasion attacks).  Adversarial example attacks are carried out during the inference phase. For this attack, the attacker doesn't change any of the model's parameters but modifies its own inputs in a way to make the model predict the inputs as the desired class.

 \blue{Recently, it has been shown that it is possible to craft adversarial examples against NIDS, as well \cite{adv_nids,sdn_adv}. Aiken et al. \cite{sdn_adv} showed one can craft adversarial examples against an NIDS deployed on an SDN for a DDoS attack by increasing packets payload size, decreasing packet rate and forging traffic with the reverse source and destination to that of the attack packets. Hashemi et al. \cite{adv_nids} have designed a more general method to craft adversarial examples for a wide range of attacks. They've shown that by applying a combination of legitimate transformations such as splitting a packet into multiple packets, changing the delay between packets, etc. one can fool the NIDS while not breaking the underlying network protocols. In this paper, in order to evaluate our NIDS in an adversarial setting, we follow their threat model and use the algorithms and transformations introduced by them to craft adversarial examples against our NIDS. More specifically, we consider that the attacker has a copy of the NIDS deployed on the victim network and knows all of its parameters. Also, we assume the NIDS deployed on the victim network receives a copy of all the packets that travel through the network entrances. Note, This can easily be achieved in an SDN with the help of the controller to make the switches forward all the network traffic to the NIDS node.}

%% file: 3_RePO.tex
\section{Anomaly Detection with Denoising Autoencoders}
\blue{In this section, we introduce our method for building an NIDS that can detect a wide range of network attacks while maintaining a low false alert level. We first show how a denoising autoencoder can be used as an anomaly detector. Then, we show how we can make its predictions more accurate and also more robust against adversarial examples. Finally, we show how it can be used as the core of an NIDS to build both packet-based and flow-based NIDS.}

\subsection{Reconstruction from Partial Observation}
\blue{As we saw in Kitsune, the reconstruction error which is calculated based on the difference between the output and input of an autoencoder can be used as a score function to detect anomalies in an unsupervised manner.
The assumption is that the model learns to reconstruct the inputs from the distribution it is trained on (e.g. benign traffic) better than inputs that come from another distribution (e.g. malicious traffic). We argue that this assumption is not necessarily true. The problem with using autoencoders and their reconstruction error in this way is that a neural network with a large enough capacity would have the capability to over-generalize and learn to rebuild the anomalous inputs, as good as the benign inputs. In other words, the score of malicious inputs will be very similar to the benign inputs making it hard to distinguish between them resulting in poor detection rate.}

\blue{Therefore, in order to solve the issue of over-generalization mentioned above, we use denoising autoencoders to force the model to solve a harder problem. We train the model in a way to reconstruct a given input based on observing some part of it. This way, the model has to not only reconstruct the visible parts of the input but also to generate the hidden parts of it. As a result, the reconstruction errors of the malicious inputs become larger than the benign inputs as the model can't reconstruct the hidden parts of the malicious inputs well enough which leads to better distinguishment between malicious and benign traffic and a higher detection rate. Also, note that as malicious inputs get further away from the decision boundary of the model (i.e. the threshold) it becomes harder to craft adversarial examples for them. Therefore, when we train the model in a way to make the gap between the malicious and benign scores larger it also becomes more robust against the adversarial example attack.}

\blue{We refer to this approach as Reconstruction from Partial Observation (RePO). More specifically, given a model $F$, we use the following loss function in our training phase:}
\[ Loss_{RePO} = \frac{1}{N}\sum_{i=0}^N || F(x_i \odot r_i) - x_i ||_2^2  \]
\noindent where $N$ is the number of inputs in the training set. $x_i$ is the i-th input sample in the training set and $r_i$ is a tensor with the same size as $x_i$. The elements in $r_i$ are randomly $1$ and $0$ and the average percentage of $0$s is $\delta$, which is a hyperparameter that should be chosen with regard to the dataset the model is trained on (we set $\delta$ to $0.75$ in all of our evaluations). In other words, we want to minimize the mean square of reconstruction errors. During inference time, we again mask some parts of the input randomly and then feed the result into the model. Finally, like any other anomaly detector we discussed so far, we need a score function. We define the score function as follows: \par
\par
\[ score(x) = \frac{1}{M}\sum_{j=1}^M|F(x \odot r)^j - x^j|^2\]
\noindent where $M$ is the number of features in $x$. $x^j$ and $F(.)^j$ are the j-th features in $x$ and its reconstructed version, respectively. Therefore, during inference time, if $score(x)$ is greater than a pre-defined $threshold$, $x$ is considered an anomaly.

\subsection{RePO+}
\blue{In our approach, since $r$ is a random matrix, by having different masks the model outputs different scores. In such a setting even if an input sample is normal, the random mask might block the most important parts of the input. Therefore, the model may not be able to reconstruct the input from what it observes. In this case, the reconstruction error would also be high. In order to solve this issue, during inference time, we replicate each sample 100 times and feed them in parallel to the model such that each of them is masked with a different mask $r_i$. We then calculate the score of all of them and group them equally into 5 groups, and from each group, we keep the minimum score. Finally, we calculate a new score by adding these 5 minimum scores together and we use this new score for deciding whether or not an input is anomalous. 
When we use 100 different masks, it becomes more likely to have "better" masks. By choosing the ones which have smaller reconstruction errors, we essentially ignore the cases that have high error because of a "bad" mask (i.e. a mask which blocks the essential features of a sample which are needed to correctly classify it). Note that, here, we don't train an ensemble of 100 different models that can take a very long time to be trained. We only train one single model that receives multiple parallel copies of each input masked with different masks during inference time. We call this approach RePO+, and we empirically found that it has a higher detection rate.}

\blue{Also, note that crafting adversarial examples against RePO+ becomes harder as for a given input the adversary should plan to modify a larger set of features to fool the model compared to when there is only one mask. This is because if only a few features get changed, they can get masked by one of the masks with a high chance which makes those changes ineffective. Moreover, since the adversary can't directly query the NIDS deployed on the victim's network (otherwise would be detected) and has to craft the adversarial example by the help of their local copy of the NIDS, Even if the adversary can fool the local copy when the sample goes through the original NIDS because of a different random mask, it would generate a different score which might be higher than the threshold and therefore will be detected. This stochastic nature of RePO+ also makes it more robust against adversarial examples. In Section~\ref{subsec:repo_perf_adv}, we empirically show the extent to which the model is more robust due to this property. }

\subsection{Building an NIDS with RePO}
\blue{Here, we discuss how to build a packet-based NIDS using the RePO technique.} \footnote{Our code is available at: https://github.com/s-mohammad-hashemi/repo} To build a packet-based NIDS with RePO, we first group the received packets by their sender and receiver IPs (i.e. the packets from A to B would be in the same group as packets from B to A). Then for each packet, we build a feature vector with the following features:
\begin{itemize}
    \item Inter-arrival time: The time between this packet and the previous packet in the group.
    \item Features extracted from Ethernet header: the length of the frame.
    \item Features extracted from IP header: IP header length, IP length, IP flags (df, mf, rb), TTL.
    \item Features extracted from TCP header: source port, destination port, sequence number, acknowledgment number, TCP flags (res, ack, cwr, ecn, fin, ns, push, reset, syn, urg), TCP window size, urgent pointer. These features would be zero if the current packet is not a TCP packet.
    \item Features extracted from UDP header: UDP length, source port, destination port. These features would be zero if the current packet is not a UDP packet.
    \item Features extracted from ICMP packet: ICMP type. This feature would be zero if the packet is not ICMP packet.
    \item Direction: This feature is a binary feature that shows whether the packet is from the IP which started the communication or from the other end.
\end{itemize}

In total, the feature vector corresponding to each packet contains 29 features. In order to make a decision for a given packet, in addition to the features of that packet, we also send the features extracted from the previous 19 packets from that group to our model. Therefore the decision is made based on observing 20 consecutive packets, and we feed 580 features to the model in each case. If there are not enough packets before a given packet, we pad it with feature vectors which all of their elements are zero. 
The model architecture we used to train RePO as a packet-based NIDS is a light-weight neural-network with only one hidden layer with 2048 neurons. 
During the training phase, we first normalized each feature separately by a min-max scaling approach. We set the batch size to be 512 and we trained the model for 30000 different batches which were selected randomly with learning rate 0.001, 0.0001 and 0.00001 each for 10000 iterations.

\blue{Note, it is not required to utilize RePO only in the packet-based context. RePO can also be used to build a flow-based NIDS when combining it with a feature extractor that extracts features at the flow level.
The model architecture we used for detecting anomalies at the flow level is a fully-connected network with 6 hidden layers each with size 256 and ReLU non-linearity in addition to 2 dropout layers after the third and fifth hidden layers of the network. During the training phase, we first normalized each feature by a min-max scaling approach. We used a batch size of 256 and trained the model for 5 epochs with a learning rate of 0.001. 
}

%% file: 4_eval.tex
\section{Evaluation}
\blue{In this section, we first briefly describe the dataset and metrics we used for our evaluation. Then, we evaluate how RePO performs in detecting network attacks in both normal and adversarial setting by comparing it against our baselines. Finally, we evaluate the system performance of our NIDS to measure the training time and its throughput in run-time.}
\subsection{Dataset}
\blue{To evaluate our approach, we used a highly cited dataset containing network traces of twelve network attacks from the Canadian Institute of Cybersecurity (CIC) \footnote{The dataset can be downloaded at: https://www.unb.ca/cic/datasets/ids-2017.html} ~\cite{cic_dataset}. These network attacks were carried out over a 5-day work week in a controlled environment and are as follows: FTP-Patator, SSH-Patator, DoS slowloris, DoS slowhttptest, DoS Hulk, DoS GoldenEye, Heartbleed, Web Attacks, Infiltration, Botnet, PortScan and DDoS. These network attacks make 10.33\% of our test set at the packet level and 24.22\% at the flow level. They are also distributed unevenly (e.g. Dos Hulk alone makes up 48\% of malicious packets). Therefore, we demonstrate the detection rate of NIDS for each attack separately.}
\blue{The whole dataset contains more than 56 million packets (2.8 million flows). We extracted features directly from the packets as described in the previous section to evaluate RePO in the packet-based context and used the flow-level features provided in this dataset to evaluate our NIDS in the flow-based context. Each flow was labeled as either benign or with the specific attack name and we labeled each packet based on the flow labels ourselves with the same procedure as Hashemi et. al in \cite{adv_nids}.
We trained the packet and flow-based NIDS on the Monday traffic, which solely contains over 11.6 million benign packets (529,481 flows). The NIDS were then tested on the network traffic generated during Tuesday-Friday, which contains both benign and network attack traffic. We excluded web attacks from our evaluations because in both packet-based and flow-based cases the features were collected from packet headers and in order to detect web attacks packet payloads should also be inspected.}

\subsection{Evaluation Metrics}
\subsubsection{True Positive Rate (TPR)} TPR shows the ratio of malicious traffic that is detected as malicious to the whole malicious traffic when the model's threshold is fixed to a specific number.
\subsubsection{False Positive Rate (FPR)} FPR shows the ratio of benign traffic that is considered as malicious to the whole benign traffic when the model's threshold is fixed to a specific number. We evaluate NIDS at a low false-positive rate to avoid the detrimental effects of base-rate fallacy~\cite{baserate}.

\subsection{Detection Performance of RePO in a Normal Setting}
\subsubsection{Packet-based NIDS:} \label{packet_based_normal_section}
\begin{figure}
  \centering
  \includegraphics[width=80mm]{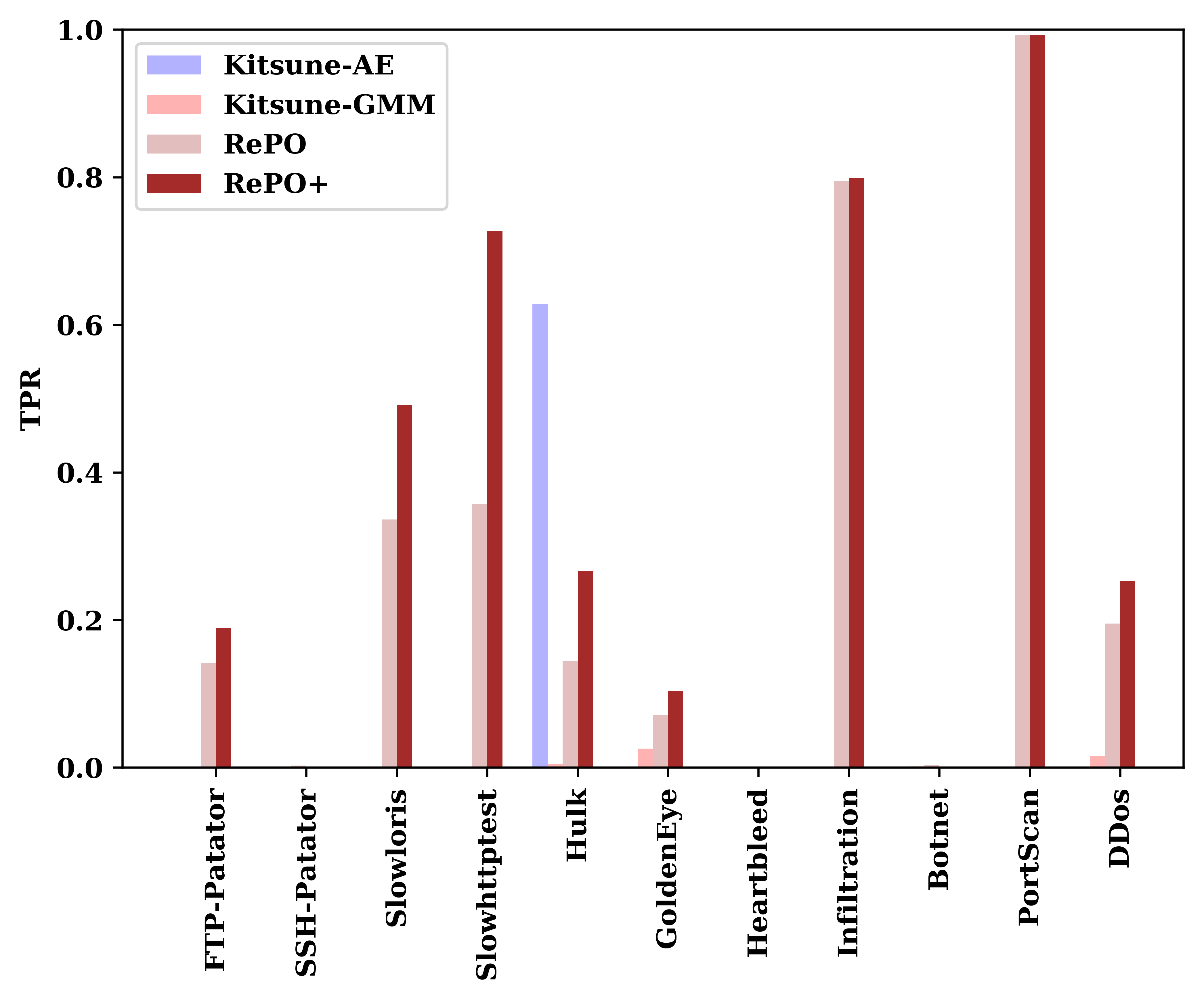}
  \caption{The TPR of packet-based NIDS for each attack when FPR is 0.01.}
  \label{fig:repo_each_attack_normal}
\end{figure}

\blue{In the packet-based context, we compare our method with Kitsune. As Mirsky et al. showed in their paper, the ensemble of autoencoders in Kitsune can also be replaced with GMM. Here, in addition to comparing with Kitsune while it uses an ensemble of autoencoders (Kitsune-AE), we compare our approach with it when it uses GMM (Kitsune-GMM), as well.
Figure \ref{fig:repo_each_attack_normal} demonstrates how our NIDS performs in detecting different attacks compared to these baselines when the threshold of each NIDS was set in a way to make FPR be 0.01. With RePO+, we can detect 8 attacks with a TPR more than FPR whereas, Kitsune-AE and Kitsune-GMM could only detect 1 and 2 attacks, respectively.}
\blue{The average detection rate of our NIDS using RePO and RePO+ across all attack categories is 27.65\% and 34.77\% while for Kitsune-AE and Kitsune-GMM it is 5.71\% and 0.44\% respectively. Thus, when using RePO+ our detection in the low false alert setting is almost 6 times better than Kitsune-AE  (29\% improvement) and 79 times better than Kitsune-GMM.}

\subsubsection{Flow-based NIDS:}

\begin{figure}
  \centering
  \includegraphics[width=80mm]{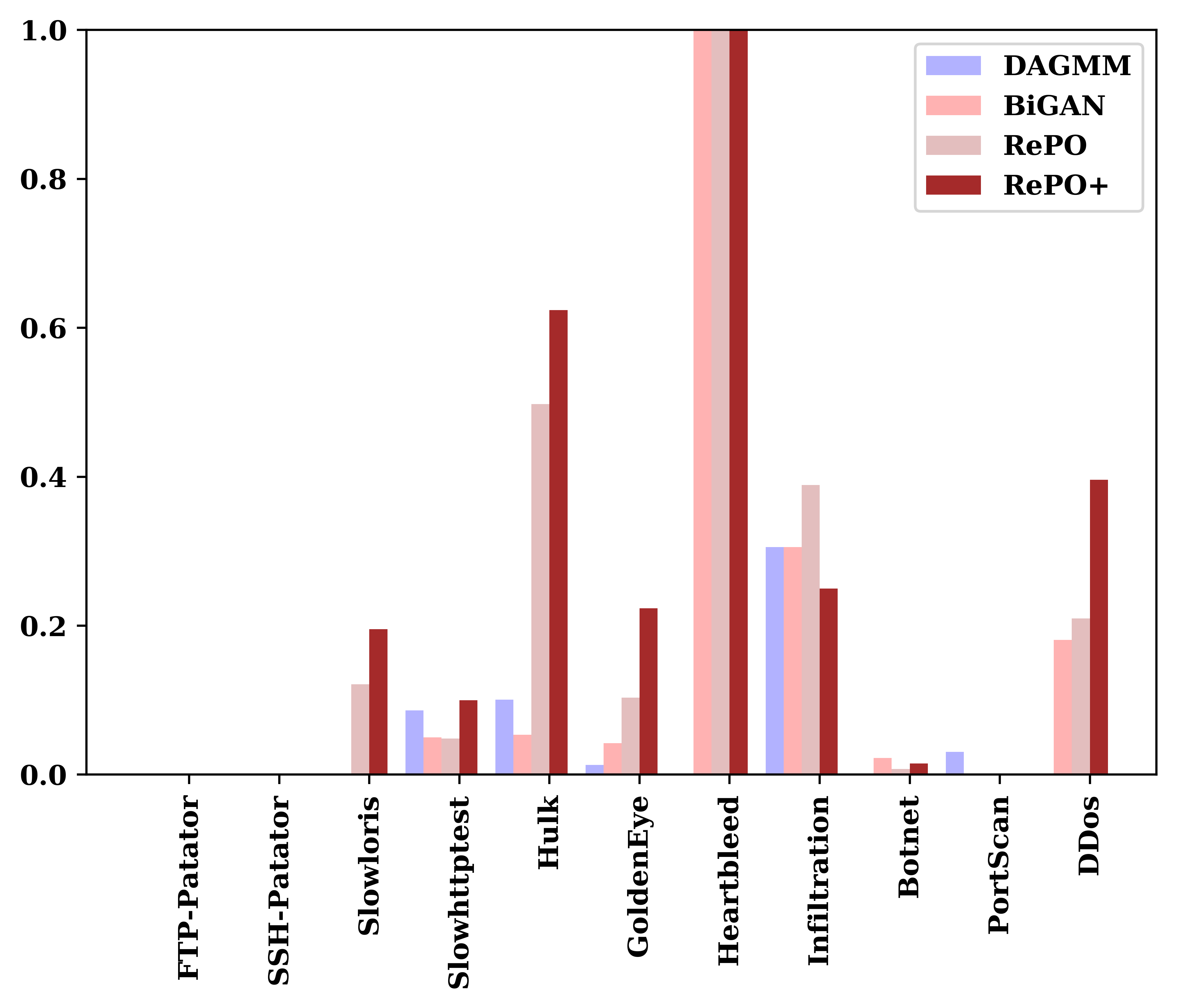}

  \caption{The TPR of flow-based NIDS for each attack when FPR is 0.01.}
  \label{fig:repo_flow_each_attack_normal}
\end{figure}

\blue{In the flow-based context, we compared RePO with the DAGMM and BiGAN-based anomaly detectors. 
Figure \ref{fig:repo_flow_each_attack_normal} demonstrates how our NIDS performs in detecting different attacks compared to these baselines when FPR is low (0.01). With RePO+, we can detect 8 attacks with a TPR more than FPR, whereas, BiGAN and DAGMM could  detect 7 and 5 attacks respectively. The average detection rate of our NIDS using RePO and RePO+ across all attack categories is 21.61\% and 25.49\% while for BiGAN and DAGMM it is 15.05\% and 4.91\% respectively. Thus, when using RePO+, our detection in the low false alert setting is 1.69 times higher than BiGAN and 5.2 times better than DAGMM.}

\blue{Finally note that there is a difference between the detection rates of packet-based NIDS and flow-based NIDS for some of the attacks such as FTP-Patator, PortScan, SlowHttpTest and Heartbleed. This is because the features extracted for packet-based NIDS are different from the features extracted for flow-based NIDS. For packet-based NIDS we only have individual values from packet headers while features for flow-based NIDS are aggregated over the whole flow and calculated differently. In addition, feature vectors we created in the packet-based scenario are extracted from packets grouped by source IP and destination IP whereas for the flow-based scenario a flow is defined based on the 5 tuples (source IP, destination IP, source port, destination port and protocol). Furthermore, some of the attacks like Botnet are hardly detected by all of the NIDS including ourselves in this low false alert setting. This is also because of the way we build our feature vectors as in both flow-based and packet-based cases the feature vectors are created based on the packets sent between a single source IP and a single destination IP. But botnet could be detected better by looking into the traffic coming from multiple source IPs. Such issues are related to feature engineering which is beyond the scope of our work and we leave designing of better feature vectors for future work. Our goal was not to detect every single network attack but to take a step forward in direction of designing an accurate and robust NIDS and showing that by using the RePO technique a better NIDS can be built in both packet-based and flow-based scenarios.}

\subsection{Detection Performance of RePO in an Adversarial Setting} \label{subsec:repo_perf_adv}
In order to evaluate our approach in an adversarial setting we use the crafting procedures introduced by Hashemi et al. in \cite{adv_nids}. Also, we follow the experiment setting introduced in \cite{adv_nids} and set the threshold of our NIDS in a way to keep FPR at 0.1. This is because our baselines didn't perform well at the low FPR (0.01) even in a normal setting and there wasn't too much malicious traffic that is detected by those NIDS at the low FPR to craft adversarial example for them. Therefore, for evaluation in an adversarial setting, we used a higher FPR (0.1) in which the robustness of different NIDS can be compared better against each other.
\subsubsection{Packet-based NIDS}

In order to evaluate the packet-based RePO+ in an adversarial setting, we tailored the crafting procedure introduced for Kitsune in \cite{adv_nids} as follows: for each packet if the packet is malicious and its score is more than the threshold, we first check whether the packet is sent from the attacker or the victim. If it was sent from the attacker, we first try to see if changing the delay between this packet and the previous packet or splitting this packet into multiple packets can fool the NIDS. If so, we make the appropriate changes and send the modified packet(s). If not, we check to see if injecting a fake TCP packet will fool the NIDS for both the fake packet and the current packet. If so, we send both of the packets; otherwise, we will only send the current packet and proceed to the next packet. If the current packet is sent from the victim, then we only check to see if injecting a packet before that works. In the case of injection, we let our algorithm bound the IAT between the current packet and the previous packet between 0 and 15 seconds (same as what considered for Kitsune). Other features can be changed between their minimum values and their maximum values. For example, in the fake packet, any of the TCP flags or IP flags can be turned on. The source port and the destination port can also be any valid value. Note that fake packets are designed in a way such that they are not processed by the victim's machine but only by the NIDS. 
Also because the crafting procedure is computationally expensive, we only applied it on the first 25000 packets of each attack.

\begin{figure}
  \centering
  \includegraphics[width=90mm]{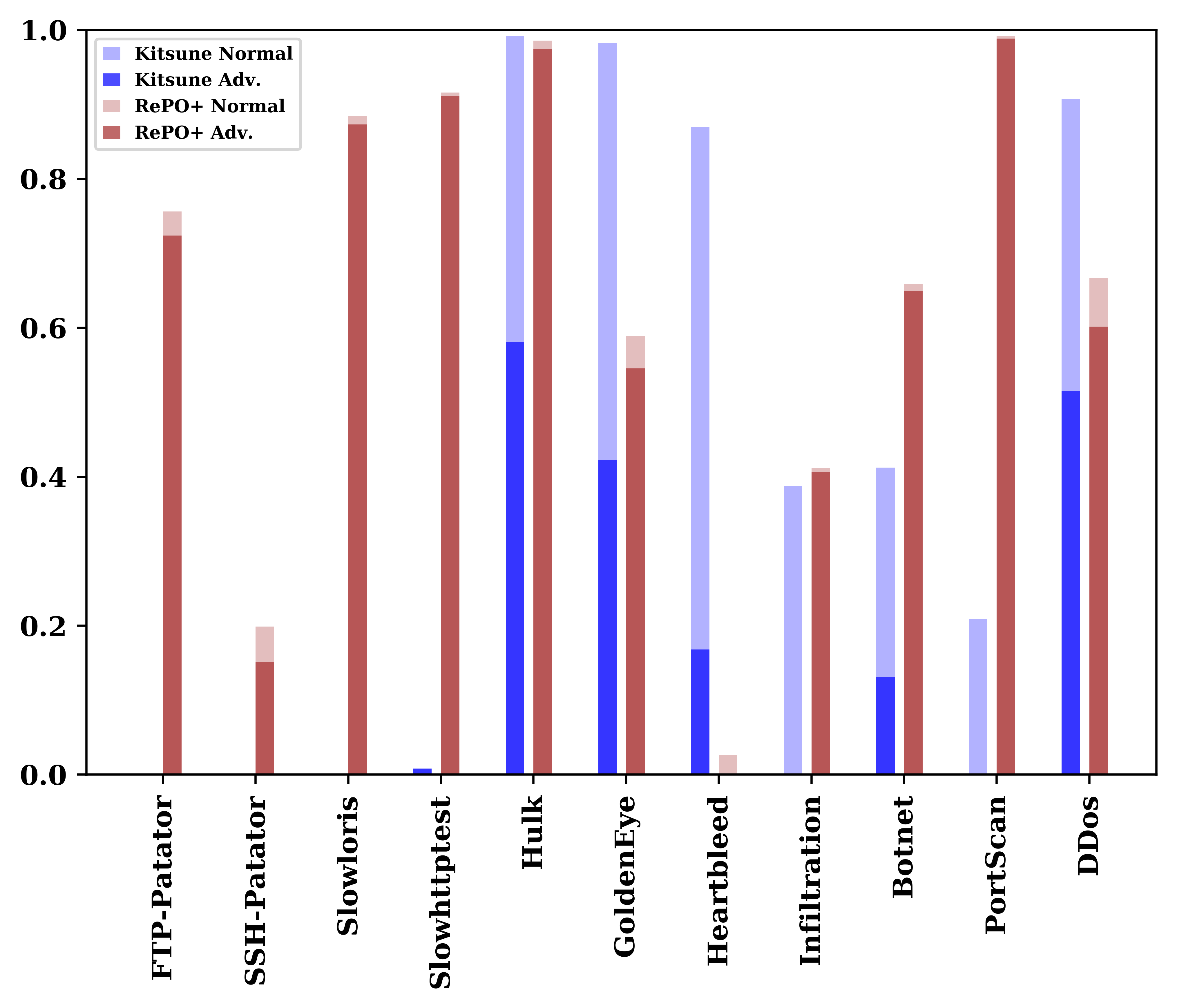}
  \caption{The TPR of RePO+ and Kitsune-GMM for each attack when FPR is 0.1 when sending normal traffic and the adversarial version of it.}
  \label{fig:REPO+_adv_ex}
\end{figure}

Figure \ref{fig:REPO+_adv_ex} demonstrates how well RePO+ performs in an adversarial setting compared to Kitsune-GMM. we only compare against Kitsune-GMM because GMM as Kitsune's detector could detect malicious traffic better than using an ensemble of autoencoders at this FPR. As can be seen, in the adversarial setting and even when accepting a higher FPR, Kitsune-GMM can only detect 5 different attacks at a rate higher than the FPR; whereas, our NIDS can detect 10 out of 11 attacks in the same setting. Also, in an adversarial setting, the average detection rate of Kitsune-GMM is 16.62\% while ours is 62.07\% (3.73x better). Kitsune's performance on average dropped 26.74\%, while our performance only dropped 2.36\%, an improvement over Kitsune of 11.33x.

\subsubsection{Flow-based NIDS}

\begin{figure}
  \centering
  \includegraphics[width=90mm]{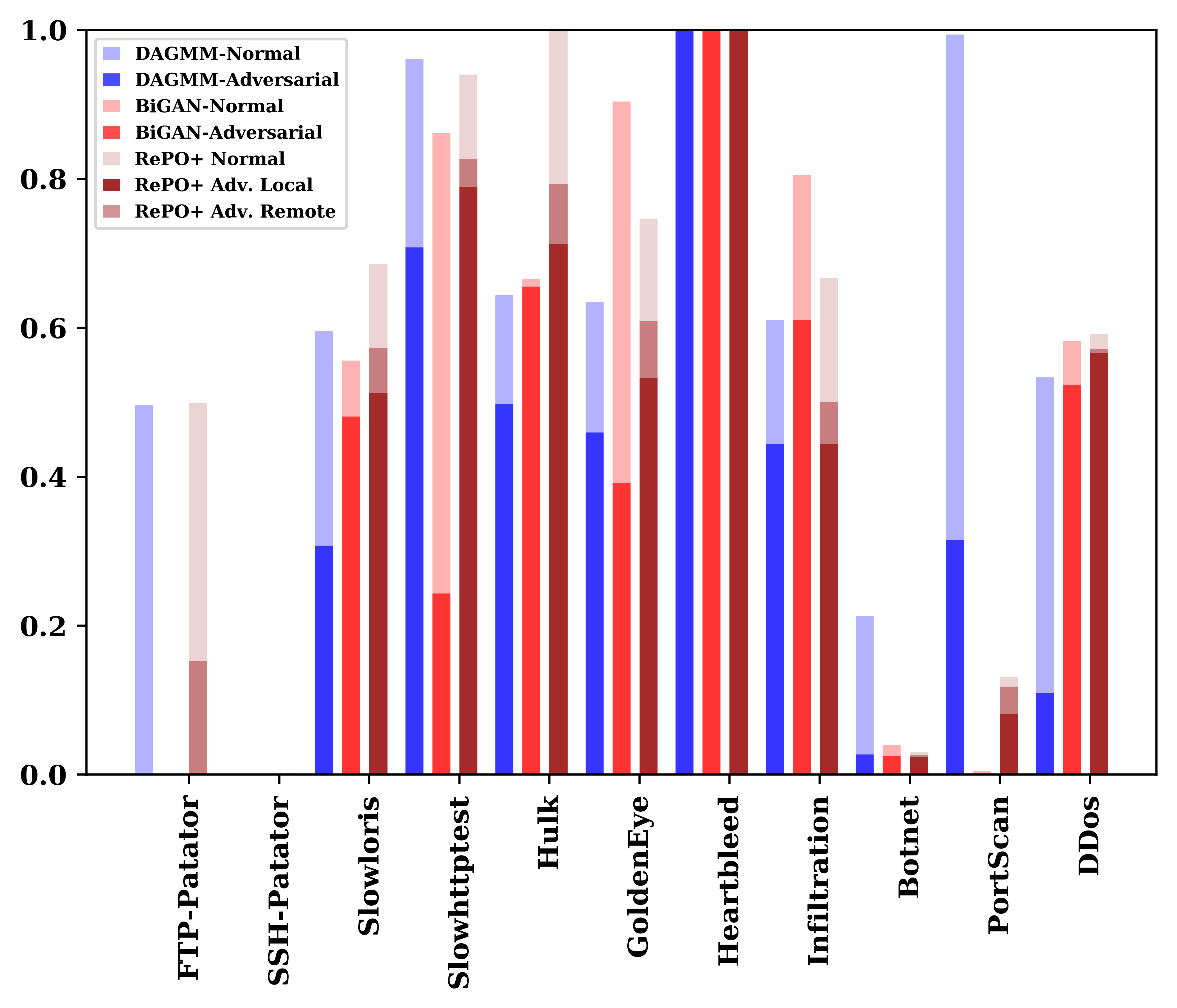}

  \caption{The TPR of RePO+, DAGMM and BiGAN for each attack when FPR is 0.1 and when sending normal traffic and the adversarial version of it.}
  \label{fig:repo_flow_each_attack_adversarial}
\end{figure}

In order to evaluate the flow-based RePO+ in an adversarial setting, we used the same procedure introduced in \cite{adv_nids} for flow-based NIDS.
Figure \ref{fig:repo_flow_each_attack_adversarial} shows how well RePO+ can detect different attacks in an adversarial setting compared to the other flow-based NIDS. The other NIDS that we mentioned had all deterministic behavior during inference time. For this reason, when the adversary crafts an adversarial version of a given flow for their own local copy, the exact same features can fool the NIDS deployed on the victim's network. But as we mentioned earlier, RePO+ predictions are not deterministic. That is to say, the RePO+ output for the same set of features might be different between the adversary's local copy and the actual NIDS deployed on the victim's site. This gives the model more robustness and the results for this case are marked with "RePO+ Adv. Remote" on the figure.
As can be seen, in an adversarial setting, and when we accept a higher rate of false alerts, BiGAN and DAGMM can detect 7 and 8 different attacks at a rate higher than FPR. Whereas, our NIDS can detect 9 out of 11 attacks in the same setting. Also, in an adversarial setting the average detection rate of BiGAN and DAGMM is 35.74\% and 35.19\%, respectively, while ours is 47.02\% (1.3x better).

\subsection{System Performance of RePO}
\blue{Our NIDS is also time-efficient. In our experiments, by using an Nvidia Titan V GPU it only took 25 seconds to train RePO+ as a flow-based NIDS. Moreover, During inference time, RePO+ could process 17,550 flows per second. That is to say, the whole test set including 2.3 million flows collected over a course of 4 days could be processed in only 131 seconds. Also, as a packet-based NIDS, we could train RePO+ in 7.5 minutes and it could process 3,285 packets per second during test time. This means that the whole test set which contains more than 44 million packets could be processed in 226 minutes. Note that, this processing rate is for when we classify every single packet which is not necessary but in section \ref{packet_based_normal_section} we did it this way to have a fair comparison against Kitsune which outputs a score for every single packet. That is to say in our evaluation we considered a window with size 20, grouped 20 consecutive packets, fed them into the model and moved the window one step forward. We can simply move this window 20 steps forward to reduce the redundancy while still considering all of the packets. In this case, RePO+ can process 65,700 packets per second making it capable of processing the whole test set in only 12 minutes. It's also worth saying that in this setting we don't sacrifice detection rate at all. The average detection rate when FPR is 0.01 is still 34.83\% which is almost the same as what reported in \ref{packet_based_normal_section}. In addition, since RePO+ can become fully parallelized we can use more GPUs to process a higher number of flows/packets per second if needed. In this case, the number of flows/packets processed per second will be multiplied by the number of GPUs used.
}

%% file: 5_conclusion.tex
\section{Conclusion}\label{sec:conclusion}
In this paper, we demonstrated how denoising autoencoders can be utilized to build a more accurate NIDS than the current state of the art methods which is also more robust against adversarial example attacks.
As we showed in our experiments, on average with our approach, we can improve the detection of different attacks by 29\% in the packet-based context and by 10\% in the flow-based context in a normal setting. Furthermore, in an adversarial setting, we can improve the detection rate by 45\% in the packet-based context and by 12\% in the flow-based context.